\documentclass[12pt]{article}
\usepackage{graphicx}
\usepackage{amssymb}
\usepackage{amsmath}
\usepackage{color}
\newtheorem{theorem}{Theorem}
\newtheorem{proposition}[theorem]{Proposition}
\newtheorem{corollary}[theorem]{Corollary}

\numberwithin{equation}{section}
\numberwithin{figure}{section}
\numberwithin{theorem}{section}
\numberwithin{teorema}{section}

\def\be{\begin{equation}}
\def\ee{\end{equation}}
\def\bc{\begin{center}}
\def\ec{\end{center}}

\title{\bf Constraints for order parameters in analogical
neural networks\footnote{\it in memoriam}}
\author{Adriano Barra\,$^1$ and Francesco Guerra\,$^2$}
\begin{document}
\date{}
\maketitle

\begin{center}
{\small
\vskip-0.5cm


\footnote{e-mail:{\tt  adriano.barra@roma1.infn.it}} Dipartimento
di Fisica, Sapienza Universit\`a di Roma \vskip-0.5cm

{and Dipartimento di Matematica, Universit\`a di Bologna}

\footnote{e-mail:{\tt  francesco.guerra@roma1.infn.it}}
Dipartimento di Fisica, Sapienza Universit\`a di Roma \vskip-0.5cm

{and Istituto Nazionale di Fisica Nucleare, Sezione di Roma1}

}
\end{center}

\vskip 0.5cm

{\small \bf------------------------------------------------------------------------------------------------------------}

\vskip-1cm
{\small

{\bf Abstract.} In this paper we study, via equilibrium
statistical mechanics, the properties of the internal energy of an
Hopfield neural network whose patterns are stored continuously
(Gaussian distributed).
\newline
The model is shown to be equivalent to a bipartite spin glass in
which one party is given by dichotomic neurons and the other party
by Gaussian spin variables. Dealing with replicated systems,
beyond the Mattis magnetization, we introduce two overlaps, one
for each party, as order parameters of the theory: The first is a
standard overlap among neural configurations on different
replicas, the second is an overlap among the Gaussian spins of
different replicas.
\newline
The aim of this work is to show the existence of constraints for
these order parameters close to ones found in many other complex
systems as spin glasses and diluted networks: we find a class of
 Ghirlanda-Guerra-like identities for both the overlaps, generalizing the well
 known constraints to the neural networks, as well as
 new identities where noise is involved explicitly.
%
%
}

\vskip-0.6cm

{\small \bf------------------------------------------------------------------------------------------------------------}


\section{Introduction}

Despite the several recent progresses in statistical mechanics of
complex systems avoiding replica trick (see for instance
\cite{broken}\cite{talaparisi}), the original Amit Gutfrund
Sompolinsky theory (AGS) \cite{amit}\cite{ags1}\cite{ags2} for
associative neural network is still nowadays lacking a complete
rigorous mathematical backbone in this sense.
\newline
In fact, while the low storage memory case \cite{amit} has been
largely understood and even some generalization considered (see
for instance
\cite{tirozzi2}\cite{tirozzi3}\cite{talahopfield1}\cite{talahopfield2}),
in the high storage limit (number of encoded memories linearly
diverging with the number of working neurons) nor the existence of
the thermodynamic limit (clearly understood for the paradigmatic
Sherrington-Kirkpatrick model \cite{limterm}) neither a complete
description of the ergodic behavior of the system
\cite{barraguerra} have been obtained yet (even though whatever
has been proved is in agreement with AGS picture obtained via the
replica trick).
\newline
While attempting progresses in finding the critical line for
ergodicity and a clear scenario for the replica symmetric regime,
in this paper we investigate the existence, for these networks, of
proper order parameter constraints, typical features of complex
systems (see for instance \cite{aizcon}\cite{guerra2}\cite{Gsum}
for discussions on linear constraints or  \cite{contgia}\cite{gg}
for higher order constraints).
\newline
In our framework, the analogical Hopfield neural network is
thought of as a bipartite spin glass in suitably defined
variables, in which the two parties interact one another via the
memory kernel.
\newline
Consequently our constraints are satisfied by the following two
overlaps: A standard overlap taking into account the similarity
among replicas at the level of the neuronal configurations (first
party), and an overlap weighting the similarity between the
Gaussian spins (second party) among different replicas.
\newline
Due to the symmetry of the interaction among the two parties, the
constraints the overlaps obey are symmetric with respect to their
permutation too: remarkably the same symmetry was already found in
the Random Overlap Structure framework \cite{ass}, where, when
analyzing the optimal structure \cite{Grost}, roughly speaking,
two mean field spin glass models were made to interact and
identities formally equivalent to our one were found
\cite{BarDes}.
\newline
The relations for these two overlaps coupled together are obtained
with standard techniques: by avoiding divergencies in the response
of the energy with respect to a change in the noise (which plays
here the role of the temperature in material systems), and as
consequences of the self-averaging of the internal energy.
\newline
Furthermore we show that the internal energy of the system can be
completely described in term of a self-overlap among the Gaussian
spins which we prove to be self-averaging $\beta$
almost-everywhere.
\newline
The paper is structured as follows: In Section $2$ the analogical
neural network is introduced and its related statistical mechanics
framework defined.
\newline
Section $3$ deals with the a detailed study of the internal
energy: it is expressed in more ways in terms of our overlaps and
the full self-average of the spin self-overlap is shown.
\newline
Section $4$ deals with its $\beta$-streaming evaluation as well as
its self-averaging properties: the whole set of identities is
proven in this section.
\newline
Section $5$ is left for outlook and conclusions.

\section{Definition of the neural network model}

The neural network model we use resembles several features of the
original AGS one: It is a mean field fully connected network such
that each neuron interacts with the whole neural community. Its
memory kernel is stored into the synaptic matrix following the
Hebb prescription \cite{hopfield} but, differently with respect to
the original AGS theory \cite{ags1}, our memory variables are not
dichotomic bit, while share the same continuous support, weighed
by a standard distribution $\mathcal{N}[0,1]$.
\newline
Concretely we introduce a large network of $N$ two-state neurons
$\pm1 \ni \sigma_i$, $i \in (1,..,N)$, which schematize the single
neuronal dynamics \cite{amit} by matching the value $-1$ with a
quiescent (or integrating) neuron and the value $+1$ with a
spiking (or firing) neuron. They interact throughout the following
 synaptic matrix $J_{ij}$ (defined accordingly the Hebb rule for
learning),
\be J_{ij} = \sum_{\mu=1}^k \xi_i^{\mu}\xi_j^{\mu}, \ee
where each random variable
$\xi^{\mu}=\{\xi_1^{\mu},..,\xi_N^{\mu}\}$ represents a pattern
already stored by the network.
\newline
As far as we deal with equilibrium properties we are marginally
concerned with the time scales involved in the dynamics, which
however are postulated to live on, at least, three different time
sectors: The spiking dynamics of each neuron, which happens on
time scales much shorter than the others involved in the
propagation of spikes trough the network, is thought effectively
as instantaneous (spin-flip). In complete opposition the synaptic
dynamics, where learning is stored into the memory kernel by
updating the synaptic matrix, happens on time scales much slower
with respect to the ones involved in the propagation of spikes
into the network and consequently the synaptic matrix is frozen at
the beginning such that no evolution for the memories is hallowed.
\newline
Between these time sectors lives the one for the equilibrium of
the neural network, that is the object of our study.
\newline
The analysis of the network assumes that the system has already
memorized $k$ patterns (no learning is investigated) and we will
be interested in the case in which this number increases
proportionally (linearly) to the system size (high storage level).
\newline
In standard literature these patters are usually taken at random
with distribution $P(\xi_i^{\mu})=
(1/2)\delta_{\xi_i^{\mu},+1}+(1/2)\delta_{\xi_i^{\mu},-1}$, while
we extend their support to be on the real axes weighted by a
Gaussian probability distribution, i.e. \be P(\xi_i^{\mu}) =
\frac{1}{\sqrt{2\pi}}e^{-(\xi_i^{\mu})^2/2}. \ee Of course,
avoiding pathological case, in the high storage level and in the
high temperature region, the results should show robustness  with
respect to the particular choice of the
 probability distribution and we should recover the standard AGS
 theory.
\newline
The Hamiltonian of the model is defined as follows \be
H_N(\sigma;\xi) = -\frac{1}{N}\sum_{\mu=1}^k
\sum_{i<j}^N\xi_i^{\mu}\xi_j^{\mu}\sigma_i\sigma_j, \ee
which, splitting the summations $\sum_{i<j}^N =
\frac{1}{2}\sum_{ij}^N - \frac12 \sum_i^N \delta_{ij}$ enable us
to write down the following partition function
\begin{eqnarray}\nonumber Z_{N,p}(\beta;\xi) &=& \sum_{\sigma}
\exp{\Big(\frac{\beta}{2N}\sum_{\mu=1}^k\sum_{ij}^N
\xi_i^{\mu}\xi_j^{\mu}\sigma_i\sigma_j -
\frac{\beta}{2N}\sum_{\mu=1}^k\sum_{i}^N (\xi_i^{{\mu}})^2 \Big)}
\\  \label{due}  &=&
\tilde{Z}_{N,p}(\beta;\xi)\exp\Big(\frac{-\beta}{2N}\sum_{\mu=1}^k\sum_{i=1}^N
(\xi_i^{{\mu}})^2\Big), \end{eqnarray}
where $\beta$, the inverse temperature in spin glass theory,
denotes the level of noise in the network and we defined
\be \tilde{Z}_{N,p}(\beta;\xi)=
\sum_{\sigma}\exp(\frac{\beta}{2N}\sum_{\mu=1}^k\sum_{ij}^N
\xi_i^{\mu}\xi_j^{\mu}\sigma_i\sigma_j ). \ee Notice that the last
term at the r.h.s. of eq. (\ref{due}) does not depend on the
particular state of the network.
\newline
Consequently we focus just on $\tilde{Z}(\beta;\xi)$. Let us apply
the Hubbard Stratonovich lemma to linearize with respect to the
bilinear quenched memories carried by the
$\xi_i^{\mu}\xi_j^{\mu}$; if we define the ``Mattis
magnetization'' \cite{amit} $m_{\mu}$ as \be\label{emme} m_{\mu}=
\frac{1}{N} \sum_i^N \xi_i^{\mu}\sigma_i, \ee we can write
 \begin{eqnarray}\label{enne} \tilde{Z}_{N,k}(\beta;\xi) &=& \sum_{\sigma}\exp(\frac{\beta
 N}{2}\sum_{\mu=1}^k
 m_{\mu}^2) \\ \nonumber &=& \sum_{\sigma}\int  \prod_{\mu=1}^k (\frac{dz_{\mu}\exp(-z^2_{\mu}/2)}{\sqrt{2\pi}}) \exp(\sqrt{\beta N}\sum_{\mu=1}^k m^{\mu}z_{\mu}).
 \end{eqnarray}
In  what follows, the following partition function, defining
implicitly an effective Hamiltonian, will be used:
\be\label{efficace} \tilde{Z}_{N,k}(\beta;\xi)= \sum_{\sigma}\int
\prod_{\mu}^k d\mu(z_{\mu})\exp\big(
\sqrt{\frac{\beta}{N}}\sum_{\mu,i}\sigma_i \xi_{i\mu}z_{\mu}
\big), \ee where $d\mu(z_{\mu})$ is the Gaussian measure.
\newline
Note that, as we have mapped the neural network problem into a
spin glass problem also the normalization factor of the effective
Hamiltonian is changed coherently.
\newline
In fact, in the high storage case, this structure clearly reflects
the interaction among the $N$ dichotomic spin $\sigma$ and the $k$
Gaussian variables $z$ through the random interaction matrix
encoded by the patterns (in the low level of stored memories,
where $N$ goes to infinity but $k$ remains finite such an
equivalence breaks down).
\newline
Reflecting this ''bipartite'' nature of the Hopfield model
expressed by eq. (\ref{efficace})  we introduce two other order
parameters
 beyond  the ``Mattis magnetization'' (eq. (\ref{emme})): the first is the standard overlap
between the replicated neurons, defined as \be q_{ab}= \frac1N
\sum_{i=1}^N \sigma_i^a \sigma_i^b  \ee and the second is the
overlap between the spins of different replicas, defined as \be
p_{ab} = \frac{1}{k} \sum_{\mu=1}^k z^a_{\mu}z^b_{\mu}. \ee

\bigskip

Taken $F$ as a generic function of the neurons, we define the
Boltzmann state $\omega_{\beta}(F)$ at a given level of noise
$\beta$ as 
\be \omega_{\beta}(F) = \omega(F)=
(\tilde{Z}_{N,k}(\beta;\xi))^{-1} \sum_{\sigma}\int
\prod_{\mu}^k d\mu(z_{\mu}) F(\sigma) \exp\big(
\sqrt{\frac{\beta}{N}}\sum_{\mu,i}\sigma_i \xi_{i\mu}z_{\mu}
\big), 
\ee 
and often we will drop the subscript $\beta$
for the sake of simplicity. The $s$-replicated Boltzmann measure
is defined as $\Omega = \omega^1\times \omega^2 \times ... \times
\omega^s$ in which all the single Boltzmann states are independent
states at the same noise level $\beta^{-1}$ and share an identical
distribution of quenched memories $\xi$.
\newline
The average over the quenched memories will be denoted by
$\mathbb{E}$ and for a generic function of these memories $F(\xi)$
 can be written as \be \mathbb{E}[F(\xi)] = \int
\prod_{\mu=1}^p \prod_{i=1}^N \frac{d
\xi_i^{\mu}e^{-\frac{(\xi_i^{\mu})^2}{2}}}{\sqrt{2\pi}}F(\xi)=
\int F(\xi)d\mu(\xi), \ee of course $ \mathbb{E}[\xi_i^{\mu}]=0$
and $ \mathbb{E}[(\xi_i^{\mu})^2]=1$.
\newline
We use the symbol $\langle . \rangle$ to mean $\langle . \rangle =
\mathbb{E}\Omega(.)$.
\newline
In the thermodynamic limit, it is assumed
$$
\lim_{N \rightarrow \infty} \frac{k}{N}= \alpha,
$$
$\alpha$ being a given real number, parameter of the theory.
\newline
The standard quantity of interest is the intensive  quenched
pressure, defined as 
\be A_{N,k}(\beta)= -\beta
f_{N,k}(\beta) = \frac{1}{N}\mathbb{E}\ln \tilde{Z}_{N,k}(\beta;\xi),
\ee where $f_{N,k}(\beta)=
u_{N,k}(\beta)-\beta^{-1}s_{N,k}(\beta)$ is the free energy
density, $u_{N,k}(\beta)$ the internal energy density and
$s_{N,k}(\beta)$ the intensive entropy.
\newline
Assuming that the thermodynamic limit of the free energy and the
internal energy exist, these quantities will be denoted as
$A(\alpha,\beta), \ u(\alpha,\beta)$.

\section{Properties of the internal energy}

In this section we study the properties of the internal energy: at
first we evaluate it explicitly in terms of our overlaps. After
showing that it can be expressed via the spin self-overlap
$\langle p_{11} \rangle$, we prove also that $\langle p_{11}
\rangle$ is completely self-averaging.
\newline
Let us start with the next
\begin{theorem}
The following expressions for the internal energy density hold in
the thermodynamic limit
\begin{eqnarray}
\lim_{N \to \infty} \frac1N \langle H_{N,k}(\sigma;\xi) \rangle
&=& \frac{\alpha}{2(1-\beta)}\Big(1-  \langle q_{12}p_{12} \rangle
 \Big), \\ \label{EinP}
 \lim_{N \to \infty} \frac1N \langle H_{N,k}(\sigma;\xi) \rangle &=&
\ \ \ \frac{\alpha}{2\beta} \ \Big( \langle p_{11} \rangle - 1
\Big).
\end{eqnarray}
\end{theorem}
\textbf{Proof}
\newline
The proof, as well as several others through the paper, uses direct
calculation and Wick theorem (see eq.(\ref{wick})) as we deal with
Gaussian distributed variables like the spins and the memories.
\newline
In fact we remember that for these quantities, considering
$f(\xi)$ as a generic well behaved function of the memories, the
following relation (integration by parts) holds: \be\label{wick} \mathbb{E} \xi f(\xi) =
\mathbb{E}
\partial_{\xi}f(\xi). \ee
So we can write
\begin{eqnarray} \langle H_{N,k}(\sigma;\xi)\rangle &=& \partial_{\beta} \frac1N
\mathbb{E}  \log \sum_{\sigma} \exp\Big(
\sqrt{\frac{\beta}{N}}\sum_{i,\mu}\xi_{i\mu}\sigma_i z_{\mu}\Big)
\\ \nonumber
&=&
\frac{1}{2N\sqrt{N\beta}}\sum_{i,\mu}\mathbb{E}\xi_{i\mu}\omega(\sigma_i
z_{\mu}) =
\frac{1}{2N\sqrt{N\beta}}\sum_{i,\mu}\mathbb{E}\partial_{\xi_{i\mu}}\omega(\sigma_i
z_{\mu}) \\ \nonumber &=& \frac{1}{2N^2}\sum_{i,\mu}\mathbb{E}
\Big( \omega(\sigma_iz_{\mu}\sigma_i z_{\mu}) - \omega(\sigma_i
z_{\mu})\omega(\sigma_i z_{\mu}) \Big)
\end{eqnarray}
which in the thermodynamic limit becomes 
\be\label{energiamista}
u(\alpha,\beta) = \frac{\alpha}{2}\Big( \langle p_{11} \rangle -
\langle p_{12}q_{12} \rangle \Big). 
\ee 
Now it is enough to show
that \be (1-\beta)\langle p_{11} \rangle + \beta \langle
p_{12}q_{12} \rangle = 1, \ee and the proof is complete.
\newline
This can be achieved as follows
\begin{eqnarray}\label{primanuova}
\langle p_{11} \rangle &\equiv& \mathbb{E}\omega(\frac1k
\sum_{\mu}z_{\mu}^2) = \frac1k\sum_{\mu} \tilde{Z}^{-1}
\sum_{\sigma}\int \prod_{\mu} dz_{\mu}
e^{-z_{\mu}^2/2}z_{\mu}^2e^{\sqrt{\frac{\beta}{N}}\sum_{i,\mu}\xi_{i\mu}\sigma_i
z_{\mu}} \\  \nonumber &=& \frac1k \tilde{Z}^{-1}\sum_{\mu}
\sum_{\sigma}\int \prod_{\mu} dz_{\mu}
(-\partial_{z_{\mu}}e^{-z_{\mu}^2/2})z_{\mu}
e^{\sqrt{\frac{\beta}{N}}\sum_{i,\mu}\xi_{i\mu}\sigma_i z_{\mu}}
\\ \nonumber &=& \frac1p\sum_{\mu}  \Big( 1  + \sqrt{\frac{\beta}{N}}\sum_i \omega(\xi_{i\mu}\sigma_i z_{\mu})
\Big) \\  \nonumber &=&  \frac1k\sum_{\mu}  \Big( 1  +
\frac{\beta}{N}\sum_i \big(
\omega(z_{\mu}^2)-\omega^2(z_{\mu}\sigma_i) \big) \Big) \\
\nonumber &=&  1 + \beta \langle p_{11} \rangle - \beta \langle
q_{12}p_{12} \rangle,
\end{eqnarray}
and the thesis is proven.  $\Box$
\newline
\newline
As we saw in eq. (\ref{EinP}), we can express the internal energy
via $\langle p_{11} \rangle$. The following theorem is therefore
important.
\begin{theorem}\label{selfaverage}
In the thermodynamic limit, $\beta$ almost everywhere, the
self-overlap $p_{11}$ completely self-averages:
\begin{eqnarray}\nonumber \lim_{N \to \infty}
\Big(\mathbb{E}\omega^2(k^{-1}\sum_{\mu}z_{\mu}^{2}) \Big) &=& \lim_{N
\to \infty} \Big(\mathbb{E}\omega(k^{-1}\sum_{\mu}z_{\mu}^{2}) \Big)^2
= \lim_{N \to
\infty}\Big(\mathbb{E}\omega(k^{-2}\sum_{\mu,\nu}z_{\mu}^2
z_{\nu}^2) \Big),\\
\label{Pselfav} \textit{i.e.}\ \ \ \langle p_{11} p_{22}\rangle &=&\langle p_{11}\rangle^{2}=\langle p_{11}^{2}\rangle.  \end{eqnarray}
\end{theorem}
\textbf{Proof}
\newline
The proof works by direct calculations and is split in two
different steps, the former linking the first two terms of
eq.(\ref{Pselfav}), the  latter linking the second with the last.
\newline
By looking at the self-averaging of the internal energy,
$$
\lim_{N \to \infty}\ \langle \big( u_{N,k}(\beta) - \langle
u_{N,k}(\beta) \rangle \big)^2 \rangle = 0,
$$
we show that $\mathbb{E}\omega^2(p_{11})=
(\mathbb{E}\omega(p_{11}))^2$ or in terms of overlaps $\langle
p_{11} \rangle^2 = \langle p_{11}p_{22} \rangle$:
\newline
Squaring both the sides of eq.(\ref{EinP}) we get
\be\label{quadrato1} \lim_{N \to
\infty}(\mathbb{E}\omega(u(N,k)(\beta)))^2  = \frac{\alpha^2}{4
\beta^2} \Big( \langle p_{11} \rangle^2 -2\langle p_{11} \rangle
+1 \Big).\ee Now we must evaluate $\mathbb{E}\omega^2(p_{11})$:
\begin{eqnarray}\nonumber
\mathbb{E}\omega^2(p_{11}) &=& \frac{k}{4\beta
N^2}\mathbb{E}\sum_{\mu \nu}\Big( \omega(z_{\mu}^2)-1 \Big)\Big(
\omega(z_{\nu}^2) - 1 \Big) \\ \nonumber &=& \frac{1}{4\beta N^2}
\mathbb{E} \sum_{\mu} \Big(
\omega^2(z_{\mu}^2)-2\omega(z_{\mu}^2)+1 \Big) \\
\label{quadrato2} &=& \frac{\alpha^2}{4\beta}\Big( \langle
p_{11}p_{22}\rangle -2 \langle p_{11} \rangle + 1\Big).
\end{eqnarray}
Subtracting eq. (\ref{quadrato2}) to eq.(\ref{quadrato1}) we
obtained the first part of eq. (\ref{Pselfav}).
\newline
To obtain the missing relation, i.e. $\langle p_{12}^2 \rangle =
\langle p_{11}p_{22} \rangle$, we must work out the
$\beta$-derivative of the internal energy. It will involve a
polynomial in the overlaps multiplied by a factor $k$. By avoiding
its $k$-divergency (as we are in the high storage memory case when
$N\to \infty$ also $k\to \infty$, linearly with $N$) we obtain the
other relation.
\begin{eqnarray}\label{fluctu}
\partial_{\beta} \langle p_{11} \rangle &=& \frac{d}{d\beta}
\mathbb{E}\frac{\sum_{\sigma}\int
d\mu(z_{\mu})(k^{-1}\sum_{\mu}z_{\mu}^2)\exp(\sqrt{\frac{\beta}{N}}\sum_{i\mu}\xi_{i\mu}\sigma_iz_{\mu})}
{\sum_{\sigma}\int
d\mu(z_{\mu})\exp(\sqrt{\frac{\beta}{N}}\sum_{i\mu}\xi_{i\mu}\sigma_iz_{\mu})}
\\ \nonumber &=& \frac{1}{2k\sqrt{N\beta}}\sum_{\mu,\nu,i}\mathbb{E}\Big(
\omega(z_{\mu}^2\xi_{i\nu}\sigma_i z_{\nu}) -
\omega(z_{\mu}^2)\omega(\xi_{i \nu}\sigma_i z_{\nu}) \Big) \\
\nonumber &=& \frac{1}{2k\sqrt{\beta N}}\sum_{\mu,\nu,i}\mathbb{E}
\partial_{\xi_{i \nu}} \Big(
\omega(z_{\mu}^2\xi_{i\nu}\sigma_i z_{\nu}) -
\omega(z_{\mu}^2)\omega(\xi_{i \nu}\sigma_i z_{\nu})  \Big) \\
\nonumber &=& \frac{1}{2k\sqrt{\beta
N}}\sum_{\mu,\nu,i}\mathbb{E}\Big( \sqrt{\frac{\beta}{N}}\big(
\omega(z_{\mu}^2 z_{\nu}^2)-\omega(z_{\mu}^2\sigma_i
z_{\nu})\omega(\sigma_i z_{\nu}) \big) - \partial_{\xi_{i
\nu}}\big( \omega(z_{\mu}^2)\omega(\sigma_i z_{\nu}) \big)\Big),
\end{eqnarray}
where
\begin{eqnarray}\nonumber
\partial_{\xi_{i\mu}} \big(
\omega(z_{\mu}^2)\omega(\sigma_i z_{\nu}) \big) &=&
\sqrt{\frac{\beta}{N}} \omega(z_{\mu}^2\sigma_i
z_{\nu})\omega(\sigma_i z_{\nu}) - \sqrt{\frac{\beta}{N}}
\omega(z_{\mu}^2)\omega^2(\sigma_i z_{\nu}) \\ \label{stream} &+&
\sqrt{\frac{\beta}{N}} \omega(z_{\mu}^2)\omega(z_{\nu}^2) -
\sqrt{\frac{\beta}{N}} \omega(z_{\mu}^2)\omega^2(\sigma_i
z_{\nu}).
\end{eqnarray}
Pasting eq. (\ref{stream}) into (\ref{fluctu}) we get
\begin{eqnarray}\nonumber
\partial_{\beta}\langle p_{11} \rangle =
\frac{1}{2kN}\sum_{\mu,\nu,i}\mathbb{E}&\Big(& \omega(z_{\mu}^2
z_{\nu}^2) - \omega(z_{\mu}^2\sigma_i z_{\nu})\omega(\sigma_i
z_{\nu}) - \omega(z_{\mu}^2\sigma_iz_{\nu})\omega(\sigma_i
z_{\nu}) \\ \nonumber &+& \omega(z_{\mu}^2)\omega^2(\sigma_i
z_{\nu}) + \omega(z_{\mu}^2)\omega(z_{\nu}^2) +
\omega(z_{\mu}^2)\omega^2(\sigma_i z_{\nu})\Big)
\end{eqnarray}
which gives \be
\partial_{\beta}\langle p_{11} \rangle = \frac{k}{2}\Big(\langle p_{11}^2 \rangle - \langle p_{11}p_{22} \rangle \Big).
\ee As we are in the high stored pattern limit ($k\to
\infty$), in the thermodynamic limit we get 
$\langle p_{11}^2 \rangle = \langle p_{11}p_{22} \rangle$, and the proof is ended. \ $\Box$
\newline
Let us call $\bar p(\beta)$ the value taken by all overlaps $p_{aa}$ in the infinite volume limit, and by $\eta_{aa}$ the rescaled fluctuations
\be \eta_{aa} = \sqrt{k}(p_{aa} - \bar{p}).\ee
Then we have the following.
\begin{corollary}
In the ergodic regime, defined by the line $\beta=
1/(1+\sqrt{\alpha})$, where the intensive free energy is given by
$A(\alpha,\beta)= \ln 2 - \frac{1}{2}\alpha \ln(1-\beta)$
\cite{amit}\cite{barraguerra}, the value of the overlap $\bar{p}$ and the
$k$-rescaled fluctuations have the following behavior
\begin{eqnarray}\label{Pergo}
\bar{p}(\beta) &=& \frac{1}{1-\beta}, \\ \label{fluz} \langle
\eta_{11}^2 \rangle &=& \frac{2}{(1-\beta)^2}.
\end{eqnarray}
\end{corollary}
\textbf{Proof}
\newline
From the relation $A(\alpha,\beta)= \ln 2 - (\alpha / 2)
\ln(1-\beta)$ we get \be \frac{\partial A(\alpha,\beta)}{\partial
\beta}= \frac{\alpha}{2}\frac{1}{1-\beta} \equiv
\frac{\alpha}{2\beta}(\bar{p} -1), \ee from which immediately we
get eq.(\ref{Pergo}).
\newline
Then we write \be
\frac{\partial \bar{p}(\beta)}{\partial \beta} =
\frac{1}{(1-\beta)^2} \equiv \frac{k}{2\beta}(\langle p_{11}^2
\rangle - \langle p_{11}p_{22} \rangle) -
\frac{1}{\beta}\bar{p}(\beta) \ee by which immediately we get
\be \frac{1}{\beta(1-\beta)^2} = \frac{1}{2\beta}\Big( \langle
\eta_{11}^2 \rangle - \langle \eta_{11}\eta_{22} \rangle \Big).\ee
Now, noticing that, at least in the ergodic region, in the
thermodynamic limit $\langle \eta_{11}\eta_{22} \rangle \to 0$, we
get the result. \ $\Box$
\newline
\newline
Note that this corollary automatically implies $\langle
q_{12}p_{12}\rangle=0$ in the ergodic regime, as it should be.

\section{Constraints}

Now we turn to the constraints: Starting with the linear
identities we state the following
\begin{proposition}\label{ACcorollary}
In the thermodynamic limit, and $\beta$ almost-everywhere, the
following generalization of the linear overlap constraints holds
for the analogical neural network
 \be\label{ACprima}
 \langle q_{12}^2p_{12}^2 \rangle - 4 \langle
q_{12}p_{12}q_{23}p_{23} \rangle + 3 \langle
q_{12}p_{12}q_{34}p_{34}\rangle  =0.\ee
\end{proposition}
\textbf{Proof}
\newline
Let us address our task by looking at the $\beta$ streaming of the
internal energy density,  once expressed via $\langle q_{12}p_{12}
\rangle$:
\begin{eqnarray}
\partial_{\beta} \langle q_{12}p_{12}\rangle &=&
\frac{1}{Nk}\sum_{\mu,i} \mathbb{E} \partial_{\beta} \omega^2
(z_{\mu}\sigma_i) = \frac{1}{Nk}\sum_{\mu,i} \mathbb{E} 2 \omega
(z_{\mu}\sigma_i)\partial_{\beta} \omega(z_{\mu}\sigma_i) \\ &=&
\frac{2}{Nk}\sum_{\mu,i}\mathbb{E}\omega(z_{\mu}\sigma_i)\xi_{i\nu}\Big(
\omega(z_{\mu}\sigma_iz_{\nu}\sigma_j)
-\omega(z_{\mu}\sigma_i)\omega(z_{\nu}\sigma_j) \Big),
\end{eqnarray}
now we use Wick theorem on $\xi$ to get
\begin{eqnarray}\nonumber
\partial_{\beta} \langle q_{12}p_{12}\rangle &=&
\frac{2}{N^2k^2}\sum_{\mu,\nu,i,j}\Big( \big(
\omega(z_{\mu}\sigma_iz_{\nu}\sigma_j)-\omega(z_{\mu}\sigma_i)\omega(z_{\nu}\sigma_j)
\big)\big( \omega(z_{\mu}\sigma_i z_{\nu}\sigma_j)-
\\ \nonumber
&+& \omega(z_{\mu}\sigma_i)\omega(z_{\nu}\sigma_j) \big) +
\omega(z_{\mu}\sigma_i)\{\omega(z_{\mu}\sigma_i\sigma_jz_{\nu}z_{\nu}\sigma_j)
- \omega(z_{\mu}\sigma_i z_{\nu}\sigma_j)\omega(z_{\nu}\sigma_j)
\\ \nonumber
&-&
\omega(z_{\mu}\sigma_i)\omega(z_{\mu}\sigma_iz_{\nu}\sigma_j)\omega(z_{\nu}\sigma_j)
+
\omega(z_{\mu}\sigma_i)\omega(z_{\nu}\sigma_j)\omega(z_{\nu}\sigma_j)\omega(z_{\mu}\sigma_i)
\\ \nonumber
&-&
\omega(z_{\mu}\sigma_i)\omega(z_{\mu}\sigma_i)\omega(z_{\nu}\sigma_j
z_{\nu}\sigma_j) +
\omega(z_{\mu}\sigma_i)\omega(z_{\nu}\sigma_j)\omega(z_{\nu}\sigma_j)\omega(z_{\mu}\sigma_i)\}
\Big).
\end{eqnarray}
Introducing the overlaps we have
\begin{eqnarray}
\partial_{\beta} \langle q_{12}p_{12}\rangle &=& k \Big( \langle
p_{12}^2q_{12}^2 \rangle - \langle p_{12}q_{12}p_{13}q_{13}
\rangle \\ \nonumber &-&  \langle p_{12}q_{12}p_{13}q_{13} \rangle
+ \langle p_{12}q_{12}p_{34}q_{34} \rangle  + \langle \bar{p}
q_{12}p_{12} - \langle p_{12}q_{12}p_{13}q_{13} \rangle
\\ \nonumber
&-&  \langle p_{12}q_{12}p_{13}q_{13} \rangle +  \langle
p_{12}q_{12}p_{34}q_{34} \rangle - \langle \bar{p}q_{12}p_{12}
\rangle +  \langle p_{12}q_{12}p_{34}q_{34} \rangle.
\end{eqnarray}
The several cancelations leave the following remaining terms \be
\partial_{\beta} \langle q_{12}p_{12}\rangle = k \Big( \langle q_{12}^2p_{12}^2 \rangle -4  \langle q_{12}p_{12}q_{23}p_{23} \rangle + 3 \langle q_{12}p_{12}q_{34}p_{34} \rangle \Big)
\ee and, again in the thermodynamic limit, in the high storage
case, the thesis is proved.  \  $\Box$

\begin{theorem}\label{GG}
In the thermodynamic limit, for almost all values of $\beta$,  the following
generalization of the quadratic Ghirlanda-Guerra relations holds
for the analogical neural network
\begin{eqnarray}
\langle q_{12}p_{12}q_{23}p_{23} \rangle = \frac{1}{2}\langle
q_{12}^2p_{12}^2 \rangle + \frac{1}{2}\langle q_{12}p_{12}
\rangle^2, \\
\langle q_{12}p_{12}q_{34}p_{34} \rangle = \frac{1}{3}\langle
q_{12}^2 p_{12}^2 \rangle + \frac{2}{3}\langle q_{12}p_{12}
\rangle^2.
\end{eqnarray}
\end{theorem}
\textbf{Proof}
\newline
Starting from
$$
 \mathbb{E}(u^2_N(\beta)) = \frac{1}{4\beta N^2 N}
 \sum_{\mu,i}\sum_{\nu,j} \xi_{i\mu}\xi_{j\nu}\omega(\sigma_i
 z_{\mu}) \omega(\sigma_j z_{\nu}),
$$
 with a calculation perfectly analogous of the one
performed in the proof of Proposition \ref{ACcorollary} we obtain
the following expression \be\label{gg1} \lim_{N \to \infty}
\mathbb{E}(u^2_N(\beta)) = \frac{\alpha^2}{4}\Big( \langle
(\bar{p}- q_{12}p_{12})^2 \rangle + 6 \langle
q_{12}p_{12}q_{34}p_{34} \rangle - 6 \langle
q_{12}p_{12}q_{23}p_{23} \rangle \Big),\ee which must be compared
with the square of the r.h.s. of eq.(\ref{energiamista}) that is
equal to \be\label{gg2} \mathbb{E}(u^2_N(\beta)) =
\frac{\alpha^2}{4}\Big( \bar{p}^2 -2 \bar{p} \langle q_{12}p_{12}
\rangle + \langle q_{12}^2 p_{12}^2 \rangle \Big).\ee As a
consequence, subtracting eq.(\ref{gg2}) to eq.(\ref{gg1}) and
taking into account also eq. (\ref{ACprima}) (that we rewrite for
simplicity) we get the linear system
\begin{eqnarray}
0 &=& \langle q_{12}^2p_{12}^2 \rangle + 6 \langle
q_{12}p_{12}q_{34}p_{34} \rangle -6 \langle
q_{12}p_{12}q_{23}p_{23} \rangle - \langle q_{12}p_{12} \rangle^2  \\
0 &=& \langle q_{12}^2p_{12}^2 \rangle -4  \langle
q_{12}p_{12}q_{23}p_{23} \rangle +3\langle
q_{12}p_{12}q_{34}p_{34} \rangle
\end{eqnarray}
whose solutions gives exactly the expressions reported in Theorem
\ref{GG}. \ $\Box$

\begin{theorem}
For the analogical neural network, a new class of identities,
which involve explicit dependence on the noise of the network,
holds in the thermodynamic limit;
\newline
examples of which are
\begin{eqnarray}\label{vecchia}
1 &=& (1-\beta)\bar{p} + \beta \langle q_{12}p_{12} \rangle, \\
\label{nuova} 0 &=& (1 + \beta \bar{p} - \bar{p})\langle q_{12}^2
\rangle + -2\beta \langle q_{12}p_{12}q_{13}^2 \rangle + \beta
\langle q_{13}^2p_{12}\rangle.
\end{eqnarray}
\end{theorem}
\textbf{Proof}
\newline
The proof of eq.(\ref{vecchia})  is simply the explicit
calculation of the quantity $\langle p_{11} \rangle =
\mathbb{E}\omega(k^{-1}\sum_{\mu} z_{\mu}^2)$, as  established in
the derivation of eq.(\ref{primanuova}) and that in the
thermodynamic limit, remembering Theorem \ref{selfaverage},
$\langle p_{11} \rangle = \bar{p}$.
\newline
The proof of eq.(\ref{nuova}) works exactly on the line of the
proof of eq.(\ref{vecchia}) by simply working out explicitly the
term $\mathbb{E}\omega(k^{-2}\sum_{\mu,\nu} z_{\mu}^2z_{\nu}^2)$
(and so on for higher order relations).  \ $\Box$

\section{Summary}

In this paper we analyzed the properties  of the internal energy
of an analogical
 neural network:
\newline
At first we mapped the problem into a bipartite spin glass and
evaluate its internal energy by introducing two order parameters
able to fulfil our task: a standard spin glass overlap comparing
neural configurations (first party) on different replicas and an
overlap among the Gaussian spins (second party) on different
replicas.
\newline
We showed that the internal energy density can be expressed via
the spin self-overlap and proved its full self-average.
\newline
Furthermore, for these overlaps, we investigate the presence of
constraints, founding both the linear  and the quadratic
identities, as expected, being the analogical neural network a
well known complex system. \newline These constraints appear with
a clear symmetric structure with respect to the two overlaps,
which interact together in both the families. Ultimately,  this
symmetry reflects the bipartite nature of the neural networks by
which interaction among the $k$ Gaussian spins and the $N$
dichotomic variables is encoded in the memory patterns $\xi$.
\newline
Future works will be developed toward the analysis of the critical
line for the ergodicity, extending previous results up to that
line \cite{barraguerra} and to the study of the still rather
obscure (at the mathematical level) retrieval of the replica
symmetric regime.

\bigskip

{\bf ACKNOWLEDGEMENTS.}
\newline
\newline
Support from MiUR (Italian Ministry of University and Research)
and INFN (Italian Institute for Nuclear Physics) is gratefully
acknowledged.
\newline
AB work is partially supported by the SmartLife Project (Ministry
Decree $13/03/2007$ n.$368$) and partially by the
 CULTAPTATION Project (European Commission
contract FP6 - 2004-NEST-PATH-043434).



\bigskip

\begin{thebibliography}{00}
\bibitem{aizcon} M. Aizenman, P. Contucci, {\em On the stability of the
    quenched state in mean field spin glass models},
  J. Stat. Phys. {\bf 92}, 765-783 (1998).

\bibitem{ass} M. Aizenman, R. Sims, S. L. Starr,
\emph{An Extended Variational Principle for the SK Spin-Glass
Model}, Phys. Rev. B, \textbf{68}, 214403 (2003).

\bibitem{amit} D.J. Amit, {\em Modeling brain function: The world of attractor neural
network} \ Cambridge Univerisity Press, 1992.

\bibitem{ags1} D.J. Amit, H. Gutfreund, H. Sompolinsky,  \textit{Spin Glass model of neural networks}, Phys.
Rev. A \textbf{32}, 1007-1018, (1985).

\bibitem{ags2} D.J. Amit, H. Gutfreund, H. Sompolinsky \textit{Storing infinite numbers of patterns in a
spin glass model of neural networks}, Phys. Rev. Lett.
\textbf{55}, 1530-1533, (1985).

\bibitem{barra1} A. Barra,
\textit{Irreducible free energy expansion and overlap locking in
mean field spin glasses}, J. Stat. Phys. \textbf{123}, 601-614
(2006).

\bibitem{bds1} A. Barra, L. De Sanctis, \textit{Overlap fluctuations from Random Overlap
Structures}, J. Math. Phys. \textbf{47},  103305  (2006).

\bibitem{BarDes} A. Barra, L. De Sanctis \textit{Stability properties and probability distribution of multi-overlaps in dilute spin glasses},
Journal of Statistical Mechanics  J. Stat. Mech.  P08025 (2007).

\bibitem{barraguerra} A. Barra, F. Guerra, \textit{About the ergodicity in Hopfield analogical neural
network}, to appear in J. Math. Phys. Special Issue "Statistical
Mechanics on Random Graphs", (2008).

\bibitem{contgia} P. Contucci, C. Giardin\`a, \textit{The Ghirlanda-Guerra
identities},  J.  Stat. Phys., \textbf{126}, N. 4/5, 917-931,
(2007).

\bibitem{gg} S. Ghirlanda, F. Guerra, {\em General properties of overlap
distributions in disordered spin systems. Towards Parisi
ultrametricity}, J. Phys. A, {\bf 31}, 9149-9155, (1998).

\bibitem{broken}  F. Guerra, {\em Broken replica symmetry bounds in the mean field spin glass model}, Comm. Math. Phys.
\textbf{233}, 1-12, (2003).

\bibitem{guerra2} F. Guerra, {\em About the overlap distribution in mean field
spin glass models}, Int. Jou. Mod. Phys. B {\bf 10}, 1675-1684,
    (1996).



\bibitem{Gsum} F. Guerra, {\em Sum rules for the free energy in the mean
field spin glass model}, Fields Institute Communications {\bf 30},
161, (2001).

\bibitem{Grost} F. Guerra, \emph{About the Cavity Fields in Mean Field Spin Glass Models},
 \texttt{cond-mat/0307673}.

\bibitem{limterm} F. Guerra, F. L. Toninelli, {\em
The Thermodynamic Limit in Mean Field Spin Glass Models}, Comm.
Math. Phys. {\bf 230}, 71-79,  (2002).

\bibitem{hopfield} J.J. Hopfield, {\em Neural networks and physical systems with emergent
collective computational abilities}, P.N.A.S. USA \textbf{79},
2554-2558, (1982).


\bibitem{tirozzi2} L. Pasteur, M. Scherbina, B. Tirozzi, {\em The
replica symmetric solution of the Hopfield model without replica
trick} J. Stat. Phys. \textbf{74}, 1161-1183, (1994).

\bibitem{tirozzi3} L.Pasteur, M. Scherbina, B. Tirozzi, {\em On the replica
 symmetric equations for the Hopfield model} J. Math. Phys. \textbf{40} 3930-3947 (1999)
\bibitem{talahopfield1}  M. Talagrand, {\em Rigourous results for the Hopfield model with many patterns},
Probab. Th. Relat. Fields \textbf{110}, 450-467, (1998).

\bibitem{talahopfield2} M. Talagrand, {\em Exponential inequalities and convergence of moments in
the replica-symmetric regime of the Hopfield model}, Ann. Probab.
\textbf{38}, 1393-1469,  (2000).

\bibitem{talaparisi} M. Talagrand, {\em The Parisi formula}, Annals of Mathematics \textbf{163},
221-263, (2006).

\end{thebibliography}
\end{document}